\documentstyle[12pt]{article} 

\begin{document}

\title{Vortex dynamics in disordered Josephson junction arrays:
from plastic flow  to  flux flow
\footnote{in Proceedings of the 
Workshop on Josephson Junction Arrays, ICTP, Trieste, Italy 
(August 1995), edited by S. R. Shenoy. 
To appear in Physica B (1996).}}    

\author{{Daniel Dom\'{\i}nguez\cite{A}}
\\
{\it Los Alamos National Laboratory, 
Theoretical Division, MS-B262,}\\
{\it Los Alamos, NM 87545 USA}}

\maketitle

\begin{abstract}
We study the dynamics of Josephson
junction arrays with positional disorder and driven by an external
current.  We consider weak magnetic fields, corresponding to a frustration
$f=n+1/25$ with $n$ integer. 
We find that above the critical current $i_c$ there
is a plastic flow of vortices, where most of the vortices are pinned
and only a few vortices flow through channels. This dynamical regime is
characterized by strong fluctuations of the total vorticity.
The number of the flow channels
grow with increasing bias current. At larger currents there is a
dynamical regime characterized by the homogeneous motion of all
the vortices, i.e. a flux flow regime. We find a dynamical phase transition
between the plastic flow and the flux flow regimes when
analyzing voltage-voltage correlation functions. 
\end{abstract}

\newpage
\section{Introduction}

The problem of transport in a medium with quenched disorder leads to
very rich physics in a great variety of systems \cite{fisher}. 
Examples include
charge density waves, Wigner solids, magnetic bubble arrays, fluid 
invasion in porous media and pinning phenomena in type II superconductors
\cite{cdw,balents,narayan,plastic,bhatta,koshelev}.
All these systems have in common a competition between 
interaction forces and randomness under an external driving force. 
For very large driving forces the randomness is believed to be
irrelevant, and the dynamics is interaction dominated. 
The system shows homogeneous flow, and all the internal
degrees of freedom move as a whole. On the other hand, for low driving
forces the dynamics is dominated by the disorder. The system shows
a very inhomogeneous flow, where the motion breaks up into pieces
moving with different velocities or not moving at all.
It has been recently discussed that a dynamical phase transition
may exist between these two kinds of flow \cite{balents,koshelev}

In the last years, a great deal of attention has been devoted to
the study of vortex pinning and dynamics in type II superconductors
as an example of this kind of problem \cite{plastic,bhatta,koshelev}.
It has been shown that for all but very weak pinning, dissipation
starts at the critical current via channels of flow of vortices.
Experiments by Bhattacharya and Higgins \cite{bhatta}
have studied the different
dynamical regimes of plastic flow and flux flow motion in superconductors.
Koshelev and Vinokur \cite{koshelev} have discussed a dynamical
melting transition from a moving vortex lattice at large currents
to a fluid flow of vortices at lower currents. 

Most of the experimental systems mentioned above have the difficulty
that there is no control of the nature and amount of disorder.
Therefore the study of disordered  Josephson junction arrays (JJA) 
\cite{jdisexp,jdis,leath}
becomes particularly promising here, since they can be specifically
fabricated with controlled randomness \cite{jdisexp}. Also, the
recent developement of imaging techniques in JJA can allow for
a direct observation of  the different spatio-temporal dynamical 
regimes expected in these systems \cite{hallen,panne}.  
In this report we consider  JJA with positional disorder \cite{jdis}
under an external magnetic field. 
In this case the effective amount of radomness
can be changed by increasing the magnetic flux per plaquette
 by integer multiples of the quantum of flux.
We have previously shown that in the limit of very large disorder
(gauge glass)
this system has a dynamical phase transition at the critical 
current \cite{yo}. Also, the dynamical phase diagrams for different amounts
of frustration and bias currents were discussed in \cite{yo2}.
In this contribution,  we consider the case when the average
number of quantum of flux per plaquette is $f=n+1/25$, with $n$
integer. In the absence of disorder, this magnetic field corresponds
to a ground state consisting in a periodic vortex lattice.
Here we will study the transition between 
the regime of plastic or inhomogeneous
flow and the regime of flux flow.

\section{Dynamical equations}

The  dynamical equations of current driven JJA  can be obtained
from considering the resistively shunted junction model for
each Josepshson junction plus current conservation at each
node \cite{acvs,yo}. One obtains the following set of dynamical
equations for the superconducting phases $\theta({\bf r})$: 
\begin{eqnarray}
\frac{d\theta (\vec r)}{dt} &=& -\frac{2\pi {\cal R}}{\Phi_0}\sum_{\vec r'} 
G(\vec r,\vec r')\{ I^{ext}(\vec r)\label{dyn}\\
&&-\sum_{\pm\hat\mu}
I_0\sin(\theta(\vec r'+\hat\mu) -\theta(\vec r')-2\pi f_{\hat\mu}(\vec r'))
\}\;,\nonumber
\end{eqnarray}
where $\theta(\vec r)$ is the phase of the superconducting wave function at
site $\vec r$, $\hat\mu=\hat x,\hat y$; $f_{\hat\mu}(\vec r)=
\frac{1}{\Phi_0}\int_{\vec r}^{\vec r+\hat\mu}
{\bf A.}d{\bf l}$ is the line integral of the vector potential with
$\Phi_0$ the flux quanta. 
The array has $N\times N$ sites, and $a$ is the lattice constant.
$I_0$ is the critical current of each junction,  assumed to be 
independent of disorder. All the
effect of the randomness is taken in $f_{\hat\mu}(\vec r)$.
 In the Landau gauge
$f(\vec r,\vec r')=-\frac{H}{\Phi_0}\frac{(r'_x-r_x)(r'_y+r_y)}{2}$, with
$H$ the applied magnetic field.
 We take random
displacements of the sites, $\vec r/a=(n_x+\delta^x,n_y+\delta^y)$,
with $n_x,n_y$ integers, and $\delta^{x,y}$ a random
uniform
number in $[-\Delta/2,\Delta/2]$. 
Noting
that $\langle \sum_{\vec R} f_{\hat\mu}(\vec r) \rangle=Ha^2/\Phi_0=f$, 
we consider here $f=n+1/25$.  The effective amount of disorder
is $W=n\Delta$ \cite{jdis}, so
one can increase the disorder by increasing $n$ (i.e. the magnetic field). 
In the absence of randomness
 ($\Delta=0$), $f=n+1/25$ is equivalent to $f=1/25$.
${\cal R}$ is the shunted resistance of each junction and 
$G(\vec r,\vec r')$ is the two dimensional lattice Green's function.
The boundary conditions are periodic along the $x$-direction.
At the
bottom (top) of the array the  external current is injected
(taken out) with $I^{ext}(r_y=0)=I$, ($I^{ext}(r_y=Na)=-I$), and 
$I^{ext}(\vec r)=0$ otherwise.
We evaluate Eq.~(\ref{dyn})  with the same fast algorithm as in
Ref.~\cite{acvs}.
The time integration is done with a
fourth order Runge-Kutta  method, with integration
step $\Delta t = 0.01-0.1\tau_J$ ($\tau_J=\frac{\Phi_0}{2\pi {\cal
R}I_0}$), for time intervals
of $t=5000\tau_J$, after a transient of $2000\tau_J$.

Magnitudes of interest are
the voltage drops along the direction of the current, 
\begin{equation}
v_y(\vec r,t)=\tau_J
\left(\frac{d\theta(\vec r+\hat y)}{dt}-\frac{d\theta(\vec r)}{dt}\right)
\end{equation}
and the distribution of vortices
\begin{equation} 
n(\vec R,t)=-\sum
\mbox{nint}\left[\frac{\theta(\vec r+\hat\mu)-\theta(\vec r)}{2\pi}
-f_{\hat\mu}(\vec r)\right],
\end{equation}
with ${\rm nint}(x)$ the nearest integer to $x$, and the sum runs around 
the plaquette $\vec R$.
We will study time averaged quantities such as the voltage
$v=\overline{\frac{1}{N_x(N_y-1)}\sum_{\vec r}
v_y(\vec r,t)}$, with $\overline{v}$ the average over time. 
 The total vorticity $n_T(t)=\frac{1}{(N-1)^2}\sum_{\vec R} n(\vec R,t)$
satisfies $\overline{n_T(t)}=f$, because of vortex 
number conservation, but it can have temporal fluctuations  
 $(\delta n_T)^2= \overline{n_T(t)^2}-\overline{n_T(t)}^2$.

\section{Transition to the flux flow regime}

If the frustration is $f=n+p/q$ and $p/q\ll 1$
there is a dilute vortex configuration in the array. 
Here we consider  $p/q=1/25$
in arrays of size $50\times 50$. In the absence of disorder ($\Delta=0$),
the vortices form a periodic lattice in the ground state.  
For intermediate values of disorder $W\sim 0.2 - 0.5$, the equilibrium
configuration of vortices at zero bias $I=0$ is a random array
of vortices, with no crystalline order. (We find that a distorted
vortex lattice exists only for very weak disorder $W<0.1$).
Here we consider the particular case of $\Delta=0.05$ and $n=4$
($W=0.2$). A full diagram of the different dynamical regimes
as a function of $W$ and bias current $I$ was given in Ref.~\cite{yo2}.
When solving numerically Eq.~(\ref{dyn}), we find that
well below a critical current $I_c$ all the vortices are pinned by
the disorder potential and there is no dissipation. 
Close but below  $I_c$ some
vortices are depinned and move during a transient time until they
are pinned again in a new stable configuration. At $I_c$
there is one channel of vortices that move without being pinned again.
Close and above $I_c$ only few vortices move (the others remain pinned)
following certain channels of flow. An example
 of this is shown in Fig.~1(a).  
As the current is increased further, more vortices
are moving and therefore more channels of flow are opened [Fig.~1(b)]. 
When increasing the current even more, eventually 
all the vortices can move  but still with
a random, fluid-like, inhomogeneous motion [Fig.~1(c)]. 
For this range of currents the voltage-current curve (see Fig.~2)
shows a nonlinear increase of the voltage $v$ from the
critical current $i_c$ (normalized by the Josephson current $i_c=I_c/I_0$). 
This regime of inhomogeneous vortex motion is usually called
{\em plastic flow}, and it
has been studied in disordered type II superconductors \cite{plastic}.
In disordered JJA, it can be 
characterized by fluctuations of the total vorticity 
$\delta n_T>0$ \cite{yo,yo2}, i.e. by the existence of flux noise.
At high currents, above a characteristic current $i_p$, above which
$\delta n_T=0$, the voltage $v$ grows linearly with the current
with an slope proportional to the vortex concentration $p/q$. This is
the so called {\em flux flow} regime. In this case all the vortices
move with almost the same velocity. 

A quantitative understanding of 
the transition from the plastic  flow to the flux flow regime can
be obtained by analyzing the correlations of the average voltages.
Let us consider the voltages along the direction of the
current, $v_y(\vec r,t)$, which are a measure of the vortex
velocities. The homogeinity of the vortex motion can be
studied from the time averaged voltages $\bar v_y(\vec r)=\bar v_y(\vec r,t)$.
We define the correlation function
\begin{equation}
\Gamma_{vx}(x) = \frac{1}{N_x(Ny-1)}
\sum_{\vec r} \bar v_y(\vec r) \bar v_y(\vec r + x\hat e_x) - v^2\,,
\end{equation}
and similarly along the $y$ direction, $\Gamma_{vy}(y)$.
In Fig.~3 we show the normalized correlation
$C_x(x)=\Gamma_{vx}(x)/\Gamma_{vx}(0)$. We obtain that the
voltage correlation can be approximated by the form
\begin{equation} 
\Gamma_{vx}(x)\sim \Gamma_{vx}(0)\,\exp(-x/\xi_{vx}).
\end{equation}
Therefore, the voltage correlation
length $\xi_{vx}$ is a measure of how inhomegeneous is the flow
along the $x$ direction. In other words, it is a measure of the typical
distance that takes a vortex
to turn from a straight path perpendicular to the current, so that
$\xi_{vx}=\infty$ (i.e. $C_{x}(x)=1$) means a perfectly straight motion.
We see in Fig.~3 that there is a dynamical phase transition at the
characteristic current $i_p$ (in this
particular case $i_p=0.405$) where $\xi_{vx}$ diverges.
We obtain that  for $i>i_p$, in the flux flow regime,
$C_{x}(x)=1$  ($\xi_{vx}=\infty$), 
i.e. all the vortices move in a straight path,
with the same velocity along the $x$ direction [this is also confirmed by 
analyzing the voltages $\bar v_x(\vec r)$]. 
On the other hand, the voltage-voltage correlations along the
$y$ direction, $\Gamma_{vy}(y)$, 
never reach complete homogeinity, therefore above $i_p$ the
voltages $\bar v_y(\vec r)$ have a dependence on $y$ only. 
This dependence consists on   small fluctuations of
$\bar v_y(\vec r)=v_y(y)$ along the $y$ direction  
around the mean value $v$.
This means that in the flux flow regime, $i > i_p$, the effect
of the disorder potential becomes irrelevant only along the
direction of the Lorentz force (the $x$ direction), but along
the $y$ direction the disorder is still important.
Another interesting result is that for low currents just
above $i_c$, the length $\xi_{vx}$ is almost constant, as it
can be seen in Fig.~3. This
corresponds with the regime in currents 
where some of the vortices flow in channels
while some other vortices remain pinned.
At high currents, of the order of the Josephson current $I_0$,
vortex-antivortices are induced in the JJA. 
A transition from the flux flow regime to dynamical regimes dominated by
the vortex-antivortex excitations appear at these
very  high currents (involving again plastic flow and homogeneos flow, but
of vortex-antivortex pairs). 
These dynamical regimes were discussed in \cite{yo2}.

\section{Discussion}

In conclusion, we have shown that the 
transition from the plastic flow regime to a flux flow regime
seems to occur as a dynamical critical phenomenon with a diverging
correlation length. 
These could be studied experimentally in JJA with controlled disorder 
at low temperatures  using
recently developed vortex imaging techniques \cite{hallen}.
Another possible test of this transition from vortex random
motion to vortex straight motion could be obtained from
measurements of the fluctuations in the Hall voltage \cite{yo3}.

\newpage
\section*{Figure Captions}

{\bf Figure 1:}
{Diagrams of vortex flow in $50\times 51$ Josephson arrays with
frustration $f=4+1/25$ and disorder $\Delta=0.05$. 
(a) $i=0.205$ ($i_c=0.203$), (b) $i=0.21$, (c) $i=0.35$,  (d) $i=0.6$
($i_p=0.405$). The black squares represent the position of the vortices
at the present time, and the gray squares the positions where the
vortices have been previously. Therefore the gray lines indicate
the paths of the vortices.}
\vspace{1.0cm}

{\bf Figure 2:}
{Voltage-current curve for a $50\times 51$ Josephson junction array
with frustration $f=4+1/25$ and disorder $\Delta=0.05$.
Currents are normalized by $I_0$ and voltages by ${\cal R}I_0$.}
\vspace{1.0cm}

{\bf Figure 3:}
{Left: plot of the voltage-voltage correlation function $C_x(x)$
for different currents. Right: plot of the correlation length $\xi_{vx}$
extracted from  $C_x(x)$ as a function of current.}

\end{document}